\documentclass[10pt,a4paper]{article}
\usepackage{t1enc}
\usepackage[latin1]{}
\usepackage[english]{babel}
\usepackage{graphicx}
\begin{document}
\begin{center}
{\bf \Large
Paraelectric-Ferroelectric Phase Transitions in Small Spherical Particles}
\end{center}

\vskip3cm
\begin{center}
O. Hudak\footnote{e-mail:hudako@mail.pvt.sk, \\
part-time work: Department of Physics, Pedagogical Faculty, Catholic University, Ruzomberok} \\
Department of Aerodynamics and Simulations, Applied Physics Group, Faculty of Aeronautics, Technical University, Kosice
\end{center}
\newpage
\section*{Abstract}

A transition in a spheroidal particle from the paraelectric to the
ferroelectric phase as well as dynamic susceptibility are studied
without approximation in the paraphase. It is assumed that the surface charge is
compensated and the boundary condition for the polarisation is
$P=0$, i. e. with zero polarisation at the surface of the particle.
There is an infinite number of resonance frequencies in the dynamic
dielectric function within the quasistatic approximation. The paraphase
properties of the dielectric response of the particle are
discussed. The transition temperature decreases with decreasing
diameter d of the particles inverse quadratically. There exists such a
critical diameter that for the particles with the diameter below
the critical one the ferroelectric phase is absent.
Comparison of the experiment with theoretical results is carried out.
Introduction of a dead layer thickness leads to a very good
agreement of theory with the experiment for $PbTiO_{3}$ and to a good
agreement with the experimentfor $BaTiO_{3}$.

\newpage
\section{Introduction}

Disordered ferroelectrics and antiferroelectrics have infrared,
microwave and Raman spectra with their characteristic features
which are not present in the ordered ferroelectrics and
antiferroelectrics. Disorder in these materials may be of a
different type. Stoichiometric order-disorder type systems are the
best studied materials of the mentioned type \cite{1}, \cite{2}
and \cite{3}. Other type of order-disorder system are those
materials in which two types of ions interchange their positions at
high temperatures. At lower temperatures there exists a freezing
of their positions. When the freezing is theoretically very slow,
this may be achieved practically by an annealing process at which the ions
order, see in \cite{2} and the references therein. The low-temperature
phase is the ferroelectric phase. A similar type of materials are
crystal-glass microcomposites. The nonstochiometric solid solutions in
which one component is a ferroelectric type material and the other
is a non-ferroelectric, or antiferroelectric type material form
another class of order-disorder systems with ferroelectric or
antiferroelectric properties. Their structural form or the form
of clusters with different dielectric properties are usually of
the nanocomposite form. New methods of the preparations of materials
have lead in the last years to a possibility to obtain nanostructural
type of the
order-disorder materials \cite{2}.

We studied the dielectric response of microcomposites of the
ferroelectric-dielectric type in our papers \cite{4} and \cite{5}.
A two-phase composite of ferroelectric-dielectric particles was
studied as concerning the dielectric response within the
quasistatic approximation. Maxwell-Garnett theory, effective medium 
theory and Bergman representation enabled us
to calculate the dielectric function for a microcomposite. The
dynamic and static properties of such composites have been reported
in these papers. There is present a distribution of polar modes
due to their presence in the original ferroelectric component in
the bulk form and due to the presence of geometrical resonances.
New low-frequency peaks appear due to the phenomenon of the geometric
resonance. Their properties are dependent on the distribution of
ferroelectric and non-ferroelectric clusters which percolate 
the material. The soft mode below the percolation transition
within a state where at least one infinite ferroelectric cluster
is present does not change its frequency depending on the
concentration of the non-ferroelectric material whatever its
strength changes. The soft mode disappears above the percolation
transition, where there is no infinite ferroelectric cluster
present. Thus the soft mode becomes a hard mode. Let us note that
presence of the multirelaxation phenomenon in the
ferroelectric-dielectric composites of the type mentioned above
has its origin in the presence of ferroelectric clusters with a
distribution of their geometrical and topological properties. In
materials of the order-disorder type with a regular crystal
structure there is a possibility to find multirelaxation
phenomenon due to other reasons \cite{6}, \cite{7} and \cite{8}
where the complex dielectric function was calculated using
semiempirical Bloch equations. A perturbation and a
non-perturbation approaches were used to calculate a dielectric
function, and the multirelaxation phenomenon was found. Its origin
is due to presence of dispersion-less excitations in ferroelectric
order-disorder materials with an incommensurate phase present at a
given temperature. Clusters with an incommensurate modulation of
the ferroelectric order parameter were observed \cite{9} not only
in dipolar glasses mentioned in \cite{3}, but also in $
Zr_{0.98}Hf_{0.02}TiO_{4} $ ceramics exhibiting an
incommensurate-commensurate phase transition due to occurence of
small polar-like regions (the domains within the incommensurate
structure surrounded by discommensurations which appear upon
cooling down from higher temperatures below some critical one). It
is interesting that there exists a similarity in response of
magnetic quantum systems with incommensurably modulated phases and
order-disorder systems with an incommensurate phase \cite{10}. In
these magnetic systems it was shown theoretically \cite{11} that in 
the quantum magnetic incommensurably modulated system exist
dispersionless excitations and that their presence may explain the 
observed Al-NQR and thermodynamic anomalies observed in $ CeAl_{2}
$ \cite{12} and \cite{13}, and in Ga-NMR study of the low-energy
excitations in $ NdGa_{2} $ \cite{14} and \cite{15}.

In dielectrics the dielectric function behaviour of a composite
depends not only on properties of infinite and finite clusters,
but also on the particles from which the composite is formed as well as
on their possible domain wall structure of the order parameter.
The size effect on the ferroelectric phase transition in $ PbTiO_{3} $
ultrafine particles was experimentally studied in \cite{16},
namely the size dependence of the transition temperature. The
difference in the critical temperature for the bulk and for the
particle with diameter d is found to be $T_{c \infty} - T_{c}(d) =
\frac{C}{(d -d_{c})}$, where the constant C is $C= 588.5^{o} $C nm
and the value for the parameter of fitting called critical
diameter $d_{c}$ of the particle $d_{c} = 12.6 $ nm for
$PbTiO_{3}$ was found. Note that it is not this critical diameter
at which the ferroelectric phase in the particle vanishes. Here
$T_{c \infty}$ is the bulk transition temperature, $T_{c}(d)$ is
the particle transition temperature. Similar effects were observed
(in earlier works) in KDP \cite{17} and in $ BaTiO_{3} $
\cite{18}.

In \cite{18.0} the surface effects on the phase transitions in 
ferroelectrics are studied using a phenomenological theory
describing the change of the local spontaneous polarisation in the
vicinity of the free surface of a ferroelectric thin film. The film
is kept between metallic electrodes. Depolarising field effects
reduce the deviation of this local polarisation from its bulk
value as compared to the surface effects on phase transitions in
other systems. The critical exponents describing the behavior of
the local polarisation in the vicinity of the Curie-Weiss
temperature $T_{c}(d)$ are the same as the bulk exponents. Only
the critical amplitudes are changed.

The Landau theory of phase transitions in thick films is developed in
\cite{18.01} for a one-component order parameter. The boundary
conditions at the surfaces of a film with thickness L are given by
means of an extrapolation length $\delta$. The exact expressions are
given for the critical temperature and the order parameters
profile in terms of elliptic functions, and the nature of the
phase transition is discussed. The dependence of the crystal structure
on the particle size in $BaTiO_{3}$ is studied in \cite{18.1}. The
Curie-Weiss temperature of $BaTiO_{3}$ powder has been
investigated in the particle size range from 0.1 to 1.0 $\mu m$.
The transformation from the ferroelectric tetragonal to the paraelectric
cubic symmetry occurs at the critical particle size of 0.12 $\mu m$
at room temperature, and the Curie-Weiss temperature drops below
the room temperature at the critical particle size. In \cite{19} the
size driven phase transition in ferroelectric particles is
studied. The spatial distribution and size dependence of the
polarisation is calculated numerically. Theoretical values for the
critical diameter $d_{c}$ in $BaTiO_{3}$ and $PbTiO_{3}$ are
compared with experimental values. The surface layer thickness
$\delta$ depends on the particle size. The predicted critical size
of the particles is smaller than that from the experiment. In
\cite{20} the size dependence of the ferroelectric transition of small
$BaTiO_{3}$ particles is studied taking into account the effect of
depolarisation. The depolarisation energy is reduced in a crystal
that contains domains of different polarisation. Authors are
considering cubic particles with alternating domains separated by
$180^{o}$ domain walls. The depolarisation energy and the
domain-wall energy contributions are taken into account in the
Landau-Ginzburg free-energy density. Assuming a hyperbolic tangent
polarisation profile across the domain wall, the domain-wall
energy and the domain-wall half thickness can be obtained by
minimizing domain free energy with respect to this thickness.
$BaTiO_{3}$ is not a perfect insulator, therefore a Schottky space
charge layer on the surface that shields the interior of the
crystal from the depolarisation field is considered. The
equilibrium polarisation P and the domain width $D_{w}$ are found. The
results show that the ferroelectric transition temperature of
small particles can be substantially lower than that of the bulk
transition temperature as a result of the depolarisation effect.
Consequently, at a temperature below the bulk transition
temperature, the dielectric constant can peak at a certain cube
size L. The results agree with the existing experimental
observations. According to the authors the theory can also be applied
to other ferroelectric materials such as $KH_{2}PO_{4}$ or
$PbTiO_{3}$. Omitting the gradient term, \cite{21}, in the free
energy expansion, the noncrystalline surface layer and
depolarisation effect lead to the critical temperature dependence
$T_{c}(d) = T_{c \infty} - \frac{6D}{\delta a_{0} d}$ where D is
connected with the correlation length $\zeta$: $D = \zeta^{2}
a_{0} \mid T - T_{c} \mid$, $a_{0}$ is the coefficient from the
Landau expansion, d is the diameter of the particle. Thus a
phenomenological study of the size effect on the phase transitions in
ferroelectric particles was done. Spontaneous polarisation
dependence on the diameter d in spherical particles is studied using
the Landau phenomenological theory. The spatial distribution of the
polarisation is obtained numerically. A size-driven phase
transition is found. In \cite{21.2} the thickness dependence of the
dielectric susceptibility of ferroelectric thin films was studied
within a phenomenological theory. If the spontaneous
polarisation is reduced in the surface layer, the mean
susceptibility of the film increases with the decrease of the film
thickness. A size-driven phase transition will take place at the
critical thickness. If the temperature-driven phase transition of
the bulk is of the second-order type, the size-driven transition will be
accompanied by a dielectric divergence. If it is the first-order, a
finite dielectric peak will appear. If the spontaneous
polarisation is enhanced in the surface layer, the mean
susceptibility of the film decreases with the decrease in the film
thickness. No size-driven phase transition and hence no dielectric
anomaly will occur in this case. In \cite{21.1} the size driven
phase transition in nanocrystalline $BaTiO_{3}$ was studied.
Nanocrystalline powders of $BaTiO_{3}$ with a narrow size
distribution were produced. The powder consisted of the
crystallite sizes whose dimensions ranged from 10 nm to 1 $\mu m$. The
paraelectric-ferroelectric phase transition was found to disappear
below a critical crystallite size of 49 nm. In accordance with
this observation a structural transition towards a cubic symmetry
became apparent in the x-ray diffractograms.

The size-induced diffuse phase transition in the nanocrystalline
ferroelectric $PbTiO_{3}$ has been studied experimentally in
\cite{21.0}. The size effects were found to become important only
below 100 nm (the coherently diffracting x-ray domain size). The
tetragonal distortion of the unit cell related to the spontaneous
polarisation decreases exponentially with size and vanishes at 7
nm. $T_{c}(d)$ decreases gradually but the transition becomes
increasingly diffuse alongside of a diminishing size from 80 to 30 nm.
The ferroelectric ordering probably persists down to 7 nm.

The size effects on cells in ferroelectric films are studied in
\cite{21.00} within the phenomenological theory. The lateral size
dependence of the Curie-Weiss temperature and the polarisation is
obtained and the stability of the ferroelectricity of the cells is
discussed. The phase-transition behavior of the spontaneous
polarisation and susceptibility of the ferroelectric thin films is
studied in \cite{21.01} that is based on the Ising model in the transverse
field. Modification of the exchange constant and the transverse
field in the surface layer might lead to the spontaneous
polarisation and the Curie-Weiss temperature changing in a
different direction as well as in the same direction according
\cite{21.01}. The Curie-Weiss temperature of the films can be
enhanced, then the temperature dependence of the spontaneous
polarisation shows a tail-like structure. There exist two peaks in
the susceptibility-temperature curve in this case. One is located
at the Curie-Weiss temperature of the films. The intensity of this
peak decreases with increasing film thickness. The other peak is
around the bulk Curie-Weiss point. The intensity of this peak
increases with increasing film thickness. The theory \cite{21.01}
gives a reasonable description of the experimental facts for the
triglycine sulfate (TGS) thin films.

Surface effects and size effects for ferroelectrics with the
first-order phase transition are studied in \cite{21.000}. The
contribution of surface to the free-energy expression was studied
for the transverse Ising model with four-spin interactions taken into
consideration. It was shown that a $P^{4}$ term should be added to
the surface terms in the free-energy expression. The surface and
size effect on polarisation and Curie-Weiss temperature were
studied using the newly developed free-energy expression.
Experimental results were discussed using the free-energy
expression. In \cite{21.001} the finite size effect in
ferroelectrics results in a structural instability. This leads to
a limitation of physical sizes and dimensions of materials in
which electric dipoles can be sustained. The size dependence on
the Curie-Weiss temperature is calculated with consideration of
crystallographic anisotropy. The mean-field theory gives that the
limitation of the transverse critical sizes is strongly dependent upon
the thickness. The limitation of the critical thickness is also
closely associated with the transverse critical sizes.

In \cite{21.5} a new phenomenological theory of size effects in
ultrafine ferroelectric particles of $PbTiO_{3}$ is discussed.
This model is taking size effects on the phenomenological
Landau-Ginzburg-Devonshire coefficients into consideration and can
successfully explain the observed size effects on the Curie-Weiss
temperature, c/a ratio, and thermal and dielectric properties of
the lead-titanate-type ferroelectric particles. Theoretical and
experimental results for $PbTiO_{3}$ fine particles are compared
and discussed. The relationship between this model and the model
of Zhong et al. \cite{21} is discussed. In \cite{21.6} the size
effects on ferroelectricity of ultrafine particles of $PbTiO_{3}$
have been studied by high resolution transmission electron microscopy.
The diameter of the ultrafine ferroelectric lead titanate particles is
ranging from 20 to 2000 nm. The crystal structure, surface
morphology, domain-wall structure and surface reconstruction have been
studied. All the particles had tetragonal structure. The lattice
constants c/a ratio and the domain size decrease with decreasing
particle size. The particles became monodomain when their diameter
was 4.2 - 20 nm. A domain wall width of 14\AA was deduced for
the $90^{o}$ domain walls.

Grain-size effects on the ferroelectric behavior of dense
nanocrystalline $BaTiO_{3}$ ceramics are studied in \cite{21.3}. A
progressive reduction of the tetragonal distortion, the heat of
transition, the Curie-Weiss temperature and the relative dielectric
constant have been studied with grain size decreasing from 1200 to
50 nm. According to the authors, the correlations between the grain
size, the tetragonal distortion, and the ferroelectric properties strongly
support the existence of an intrinsic size effect. From the
experimental trends the critical size for disappearance of
ferroelectricity has been evaluated to be from 10 to 30 nm. The
strong depression of the relative permittivity observed for the
nanocrystalline ceramics is ascribed by the authors to the combination
of the intrinsic size effect and to the size-dependent dilution
effect of a grain boundary dead layer. Size effects on Curie-Weiss
temperature of ferroelectric particles were studied in \cite{21.4}
within a model without any free adjustable parameters. The model
predicts that $T_{c}$ decreases with decreasing particle size. The
predictions of the model are in agreement with experimental
results for $PbTiO_{3}$ and $BaTiO_{3}$. For similar studies see
also \cite{22}.

Influence of the domain wall structure on the dielectric response
in $ KH_{2}PO_{4} $ type crystals was studied in \cite{23}, where
one can find also references to domain wall dynamics effects in
KDP and $ CsH_{2}AsO_{4} $. In this type of the materials
the domain walls freeze-in below some temperature and a
multirelaxation behaviour of the dielectric function is observed.
The Landau theory of $ 180^{o} $ domain walls in $ BaTiO_{3} $
type ferroelectric particles was studied recently in \cite{24} for
the particles of the rectangular form. It was assumed that the surface
charge is completely compensated. The experimentally found temperature
dependence on the size of the particle is more precisely described
by a domain wall if the sixth-order
term in the Landau free-energy expansion is taken into account
\cite{24}. Dielectric
constant and correlation length of ferroelectric particles in the
paraelectric phase was studied in \cite{24.1}. The finite size of
a particle influences the temperature dependence of the dielectric
constant and the correlation length. The deviation from the
Curie-Weiss law in the ferroelectric particles originates from the
decrease of the long-range correlation. Glinchuk et al. \cite{26}
studied variationaly the depolarisation and the surface tension
effects for polarization, susceptibility and critical temperature
$T_{c}(d)$ diameter dependence for spherical particles.

In this paper we are studying other
sources of the observed complicated multirelaxation behaviour of
the dynamic susceptibility in ferroelectric-dielectric composites
by analytical calculations.
Phase transition in a spheroidal particle from the paraelectric to
the ferroelectric phase as well as a dynamic susceptibility are
affected by the spheroidal (more generally by ellipsoidal) shape
of the ferroelectric particles. Assuming that the surface charge had
been compensated,
the spheroidal shape does not lead to plane polarisation waves due
to the surface effect: the boundary condition in a spheroidal
particle with a surface layer lead to a multirelaxation response.
There is an infinite number of responding modes, thus there is
also an infinite number of resonance frequencies. It is the aim of
this paper to discuss namely these effects of the shape of a
ferroelectric particle on its dielectric response. Firstly we
describe a model which enables us to calculate a dynamic dielectric
function within quasistatic approximation. Deriving dynamic
equations for the order parameter, the electric polarisation vector
components, we discuss the paraphase properties of the response. As a
result we find multirelaxation behaviour of the dielectric
susceptibility in the real representation. It is then easy to find
frequency dependent form of the dielectric susceptibility of a
spheroidal particle. Its form enables us to find resonance
frequencies, their number is infinite. Consequently the transition
temperature dependence on the particle diameter and on other
parameters of the model may be found easily and compared with
the experimental results.

\section{Model}

Let x be a generalized dipole length of the normal polarisation
mode with an effective charge q. The one-component order
parameter P, electric polarisation component, has the form P=xq. Let m be an effective mass of
the mode. The relaxation equation has the form which follows from
the free energy F for the systems of the fourth order in
the order parameter P, see in \cite{19}:

\begin{equation}
\label{1}
F = \int dv [\frac{A}{2} P^{2} + \frac{B}{4} P^{4} + \frac{D}{2}( \nabla P)^{2} ] + \int ds [\frac{D}{2} \frac{P^{2}}{\delta}] - \int dv [E.P]
\end{equation}

The first part of the equation (\ref{1}) is a volume contribution
to the free energy, the second one is a surface contribution to
the free energy. Last term is due to the applied electric field E
acting on the electric polarisation P. Here A, B, and D are usual
parameters of the Landau free energy expansion, E is an electric
field, $ \delta $ is a surface parameter defined by, see in
\cite{21}:

\begin{equation}
\label{2}
\frac{1}{\delta} = \frac{5J - 4J_{s}}{a_{0}J}
\end{equation}

where J is a ferroelectric interaction constant in the bulk, 
$ J_{s} $ is a ferroelectric interaction constant on the surface,
$ a_{0} $ is the bulk lattice constant. In (\ref{2}) it is assumed
for simplicity that the bulk lattice structure is a cubic one.
Note that the limit $ \delta = 0 $ corresponds to strong (anti-)
ferroelectric interactions on the surface of the particle. On the
other hand for strong bulk ferro- (antiferro-) interactions J the
constant $ \delta $ remains finite and nonzero. For the ferroelectric
particles with $ J = J_{s} $ the constant $ \delta $ is
independent on the characteristic size of the film. In spherical
particles the constant $ \delta $ is dependent on the diameter d
of the particle:

\begin{equation}
\label{2.1}
\frac{1}{\delta} = \frac{5}{d} + \frac{1}{\delta_{m}} [1 - \frac{a_{0}}{d}]
\end{equation}

where $ \delta_{m} $ is the extrapolation length for $ d = \infty
$. As noted in \cite{21}, to get the spatial distribution of the
polarisation it is necessary to minimise the free energy (\ref{1})
for a given boundary conditions which is difficult to solve
analytically. The authors of \cite{21} found the spatial
distribution of polarisation numerically. We have found this
distribution analytically for very large values of the ratio $
\frac{D}{\delta} $, for which the surface contribution to the free
energy (\ref{1}) is minimised by zero surface polarisation.

\section{Equations for the Soft Mode}

The dynamic equation for the generalized coordinate x has the form:

\begin{equation}
\label{3}
m \frac{\partial^{2}}{\partial t^{2}} x + \Gamma \frac{\partial}{\partial t} x = -q (\frac{\delta F}{\delta P} -
\frac{\partial}{\partial r} \frac{\delta F}{\delta P} ) = q E_{eff}
\end{equation}

where $ E_{eff} $ is an effective electric field acting on the
charge q, here $P = q.x$ is the polarisation dependent in the
position r in the particle. Using the free energy F form from
(\ref{1}) we find that the equation of motion (\ref{3}) has the
form:

\begin{equation}
\label{4}
m \frac{\partial^{2}}{\partial t^{2}} x + \Gamma \frac{\partial}{\partial t} x = qE + q(-AP - BP^{3} + D( \frac{\partial^{2}P}{\partial r^{2}}  + \frac{2}{r} \frac{\partial P}{\partial r}))
\end{equation}

It is convenient to introduce the mass $ m^{*} $ related to the charge q
and the relaxation constant $ \Gamma $ related to the charge q:

\begin{equation}
\label{5}
\Gamma^{*}= \frac{\Gamma}{q^{2}}
\end{equation}
\[ m^{*}= \frac{m}{q^{2}} \]

The equation of motion (\ref{4}) has now the form using (\ref{5}):

\begin{equation}
\label{6}
m^{*} \frac{\partial^{2}}{\partial t^{2}} P + \Gamma^{*} \frac{\partial}{\partial t} P = E -AP - BP^{3} + D( \frac{\partial^{2}P}{\partial r^{2}}  + \frac{2}{r} \frac{\partial P}{\partial r} )
\end{equation}

The form (\ref{6}) of the equation of motion is the form which we
will use in next sections. The boundary condition for a particle
with the diameter d has the form \cite{21}:

\begin{equation}
\label{7}
\frac{\partial P}{\partial r} +  \frac{P}{\delta } = 0
\end{equation}

at the surface $ r = \frac{d}{2} $. It is further assumed that the
polarisation P(r) in (\ref{6}) remains finite at the centre of the
particle $ \lim_{r \rightarrow 0} P(r) < \infty $. Note, that the
equation (\ref{7}) is an algebraic equation relating the P value
at the point $ r = \frac{d}{2} $, and the $ \frac{\partial
P}{\partial r} $ value at the same point. In the limit $
\frac{\delta}{D} = 0 $, the surface boundary condition is $ P(r =
\frac{d}{2}) = 0 $. This boundary condition will be used in this
paper. This condition is also compatible with existence of the
surface layer with zero electric polarization (dead layer) which
may exist around the polarized region of the particle.

\section{Polarisation Mode in the Paraelectric Phase}

The fourth-order term in the free energy (\ref{1}) may be
neglected in the mode analysis if the temperature T is above the
Curie-Weiss temperature $T_{c}$ at which the paraelectric phase
transforms to the ferroelectric one. In our model above, the free
energy corresponds to a displacive type of ferroelectrics which
undergo a second order phase transition described by a one
dimensional order parameter P. The order-disorder case has to be
treated in another way.

In the free energy F the constant A is of an obvious form $ A =
a_{1}(T - T_{c}) $, with the positive constant $ a_{1} > 0 $. Let
us assume that our particle of the spherical shape is under
influence of the time dependent spatially homogeneous electric
field $ E = E_{\omega}. \exp(j\omega t) $ oscillating with the
frequency $ \omega $, here j is an imaginary unit. Thus we assume
that the wavelength of the electric field is much larger than the
particle characteristic size. Then electric polarisation $ P =
P(r). \exp(j\omega t) $ has an amplitude, complex in general,
which is dependent on the position in the particle, and its time
dependence is of the same type as of the external electric field.
The dynamic equation of motion (\ref{6}) has the form of a second
order differential equation for the polarisation $P(r)$:

\begin{equation}
\label{8}
(- m^{*} \omega^{2} + j \Gamma^{*} \omega + A) P(r) = E_{\omega} + D( \frac{\partial^{2}P}{\partial r^{2}}  + \frac{2}{r} \frac{\partial P}{\partial r} )
\end{equation}

Substituting $ P(r) = \frac{1}{\sqrt{r}} u(r) $ into the equation
(\ref{8}), where $ u(r) $ is an unknown function, we find the
following form of the equation of motion (\ref{8}) for $ u(r) $ :

\begin{equation}
\label{9}
(- m^{*} \omega^{2} + j \Gamma^{*} \omega + A) u(z) = E_{\omega} (\sqrt{\frac{d}{2}})\sqrt{z} +
\frac{D}{(\frac{d}{2})^{2}}(\frac{\partial^{2}u}{\partial r^{2}} +
\frac{1}{z} \frac{\partial }{\partial r}u(z) - \frac{(\frac{1}{2})^{2}}{z^{2}} u(z) )
\end{equation}

Here $ r = z \frac{d}{2} $ where the variable z, $ 0 \leq z \leq 1
$, was introduced. The corresponding boundary condition has a
simple form. At the centre of the sphere the polarisation remains
finite $ \lim_{r \rightarrow 0} \frac{u(r)}{\sqrt{r}} < \infty $.
At the surface we have $ u (z=1) = 0$. To solve the equation of
motion (\ref{9}) let us start from the well-known Bessel equation of
the form \cite{27} and \cite{28}:

\begin{equation}
\label{10}
\frac{\partial^{2}u}{\partial r^{2}} + \frac{1}{z}\frac{\partial }{\partial r}u(z) - \frac{(\frac{1}{2})^{2}}{z^{2}} u(z) = - b^{2} u(z)
\end{equation}

The equation (\ref{10}) has as its solution the Bessel functions $
J_{\pm \frac{1}{2}} (\lambda_{i} z) $, where $ b^{2} \equiv
\lambda^{2}_{i} $. The root $ \lambda_{i} $ is defined as that
point for which the Bessel function vanishes $ J_{\pm \frac{1}{2}}
(\lambda_{i}) = 0 $.

The equation of the motion (\ref{9}) for the order parameter P(r) in
its transformed form $ P(r) = \frac{1}{\sqrt{r}} u(r) $ has the
solution $ u(r) = \sum_{i=1}^{i=\infty} a_{i}
J_{+\frac{1}{2}}(\lambda_{i} z) $ where z is expressed through r
as it is given above, which is finite in the centre of the sphere
$ r = 0 $ due to the fact that the relevant Bessel functions $
J_{+\frac{1}{2}}(\lambda_{i} z) $ in the corresponding limit are
finite for every i, $ \lim_{z \rightarrow 0}
\frac{J_{+\frac{1}{2}}(\lambda_{i} z)}{\sqrt{z}} =
\sqrt{\frac{2\lambda_{i}}{\pi}}$. Note that one finds $ \lim_{z
\rightarrow 0} \frac{J_{-\frac{1}{2}(\lambda_{i} z)}}{\sqrt{z}} =
\infty$, and thus the Bessel functions $
J_{-\frac{1}{2}}(\lambda_{i} z) $ do not satisfy the condition of
the finite solution in the centre of the sphere. It is well known,
that the roots for the Bessel function $
J_{+\frac{1}{2}}(\lambda_{i} ) = 0 $ are $ \lambda_{i} = i \pi $.
Expanding in (\ref{9}) also the square root of z into Bessel
functions:

\begin{equation}
\label{11}
\sqrt{z} =\sum_{i=1}^{i=\infty} b_{i} J_{+\frac{1}{2}}(\lambda_{i} z) ,
\end{equation}

where $ b_{i} = - \frac{\sqrt{2 \pi}}{\sqrt{\lambda_{i}}} (-1)^{i} $,
we find that the expansion coefficients $ a_{i} $ have the form:

\begin{equation}
\label{12}
a_{i} = \frac{- E_{\omega} \sqrt{\frac{d}{2}} (\frac{\sqrt{2 \pi}}{\sqrt{\lambda_{i}}}(-1)^{i})}{- m^{*} \omega^{2} + j \Gamma^{*} \omega + A + \frac{4 \pi^{2} D i^{2}}{d^{2}}}
\end{equation}

The space dependence of the electric polarisation P(r) is given by:

\begin{equation}
\label{13}
P(r) = \chi(r, \omega) E_{\omega}
\end{equation}

concerning the amplitude of the polarisation. Here $ \chi(r,
\omega) $ is the r-dependent susceptibility:

\begin{equation}
\label{14}
\chi(r,\omega) = \frac{1}{r} \sum_{i=1}^{i=\infty} \frac{ \frac{d}{\pi i}(-1)^{i+1}\sin(\frac{2 \pi r i}{d})}{- m^{*} \omega^{2} + j \Gamma^{*} \omega + A + \frac{4 \pi^{2} D i^{2}}{d^{2}}}
\end{equation}

which is of a multirelaxation type. As we see, in (\ref{14}) there
is an infinite number of the oscillator-like contributions with the
corresponding damping, and there exists a phase shift between the
mode contributions, which is r-dependent. Note, that at the centre
of the particle the susceptibility remains finite.

\section{Dynamic and Static Susceptibility}

The dynamic susceptibility $ \chi(\omega) $ for the particle as a
whole is that susceptibility, which we obtain from (\ref{14}) by
summing up the contributions over different positions within the
sphere, relating this quantity to a unit volume of the particle:

\begin{equation}
\label{15}
\chi(\omega) = \frac{1}{V}\int_{0}^{\frac{d}{2}} \chi(r, \omega) 4\pi r^{2} dr =
\sum_{i=1}^{i=\infty} \frac{\frac{6}{(\pi i)^{2}}}{m^{*} (\omega^{2}_{i} - \omega^{2})
+ j \Gamma^{*} \omega }
\end{equation}

Here $ V = \frac{\pi d^{3}}{6} $ is the volume of the particle.
We see that the strength of every oscillator contribution decreases
in the power law, i.e. as $i^{-2}$, for i (an integer) increasing.

The oscillator resonance frequency $ \omega_{i} $ is given by:

\begin{equation}
\label{16}
\omega_{i}^{2} = \frac{1}{m^{*}} (A + \frac{4 \pi^{2} i^{2} D}{d^{2}})
\end{equation}

We see that different resonance frequencies increase as $ i^{2} $ , and
the last term in (\ref{16}) decreases as $ d^{2} $ with increasing diameter d.
The relaxation constant remains unchanged.

Static susceptibility may be easily found from (\ref{15}) and has the form:

\begin{equation}
\label{17}
\chi_{stat} = \frac{1}{V}\int_{0}^{\frac{d}{2}} \chi(r, \omega = 0) 4\pi r^{2} dr =
\sum_{i=1}^{i=\infty} \frac{\frac{6}{(\pi i )^{2}} }{m^{*} \omega^{2}_{i} }
\end{equation}

The limits of zero frequency and of volume average are
interchangeable limiting processes due to the finite volume of the
particle. It can be easily seen from (\ref{17}) that the static
susceptibility is no more of the simple Curie-Weiss type which
would correspond to the first term in the sum.

The critical transition temperature $T_{c}(d)$ can be found from
the lowest frequency mode, the soft mode, and it is given by:

\begin{equation}
\label{18}
T_{c}(d) = T_{c, \infty} - \frac{4 \pi^{2} D}{a_{1} d^{2}}
\end{equation}

It increases to its bulk value $T_{c, \infty}$ quadratically with
the increase of the particle diameter d. There is a size driven phase
transition with a decreasing diameter d. For the diameter lower than the
critical diameter $d \leq d_{c} \equiv 2 \pi\sqrt{
\frac{D}{a_{1}T_{c, \infty}}}$ the ferroelectric phase vanishes.

\section{Comparison with Experiments}

We have compared the theoretical dependence (\ref{18}) found in
this paper for the critical temperature dependence on the diameter d
of the particle with experimental dependence of the critical
temperature on the particle diameter for $PbTiO_{3}$ and
$BaTiO_{3}$ particles. Experimental results for $PbTiO_{3}$ are
found from the measurements in \cite{16}. The authors of \cite{16}
used a theoretical formula $T_{c \infty} - T_{c}(d) = \frac{C}{(d
-d_{c})}$, where they have found the constant C, $C= 588.5^{o} $C
nm, and the value for the critical diameter d of the particle
$d_{c} = 12.6 $ nm to fit the experimental points. Thus they have
found that there exists a critical diameter below which there is
no ferroelectricity. Experimental results for $BaTiO_{3}$ are
based on the measurements in \cite{18} and used in \cite{21}. They
were described by the theoretical formula $T_{c}(d) = T_{c \infty}
- \frac{6D^{,}}{\delta A d}$ where $D^{,}$ is connected with the
correlation length $\zeta$: $D^{,} = \zeta^{2} a_{1} \mid T -
T_{c} \mid$, $a_{1}$ is the coefficient from the Landau expansion,
d is the diameter of particle. The critical diameter was found to
be 44 nm theoretically and 115 nm experimentally for $BaTiO_{3}$.
The same authors have found that for $PbTiO_{3}$ the value is 4.2
nm (theoretical value) and 13.8 nm (experimental value from
\cite{16}) and 9.1 nm (experimental value from \cite{21}). It is
interesting that Glinchuk et al. \cite{26} obtained a formula for
the critical temperature $T_{c}(d)$ dependence on the diameter of a
particle by a variational method which has the form $T_{c}(d) =
T_{c, \infty}(1-\frac{d_{L}}{d}-\frac{d_{\lambda}}{d-d_{c}})$
where $d_{L}$ is the correlation diameter renormalized by the
surface tension, $d_{c}$ is the correlation diameter renormalized
by the surface tension and by the depolarisation field and
$d_{\lambda}$ is a parameter. As we can see, the surface tensions and
depolarisation fields lead to the dependences of the critical
temperature used in the papers above as theoretical formulas for
fitting the experimental points. Our calculations in this paper of the
critical temperature dependence on the particle diameter are
exact. It is known that variational calculations always lead to a
higher free energy than exact calculations, thus our formula
(\ref{18}) (depolarisation field is assumed to be zero) which
leads to $\frac{1}{d^{2}}$ dependence instead the Glinchuk et al
$\frac{1}{d}$ dependence, is expected to be better for neglected
depolarisation effects. If the depolarisation effects
are taken into account we may expect that a constant $d_{m}$ in
our formula (\ref{18}) will be introduced in
$\frac{1}{(d-d_{m})^{2}}$ instead of $\frac{1}{d^{2}}$, as a
similar dependence is found in the Glinchuk et al. formula. This
leads to the following fitting formula for the critical transition
temperature $T_{c}(d)$:

\begin{equation}
\label{19}
T_{c}(d) = T_{c, \infty}(1 - \frac{d_{o}^{2}}{(d-d_{m})^{2}})
\end{equation}

\begin{figure}[b]
    \centering
        \includegraphics[width=1.00\textwidth]{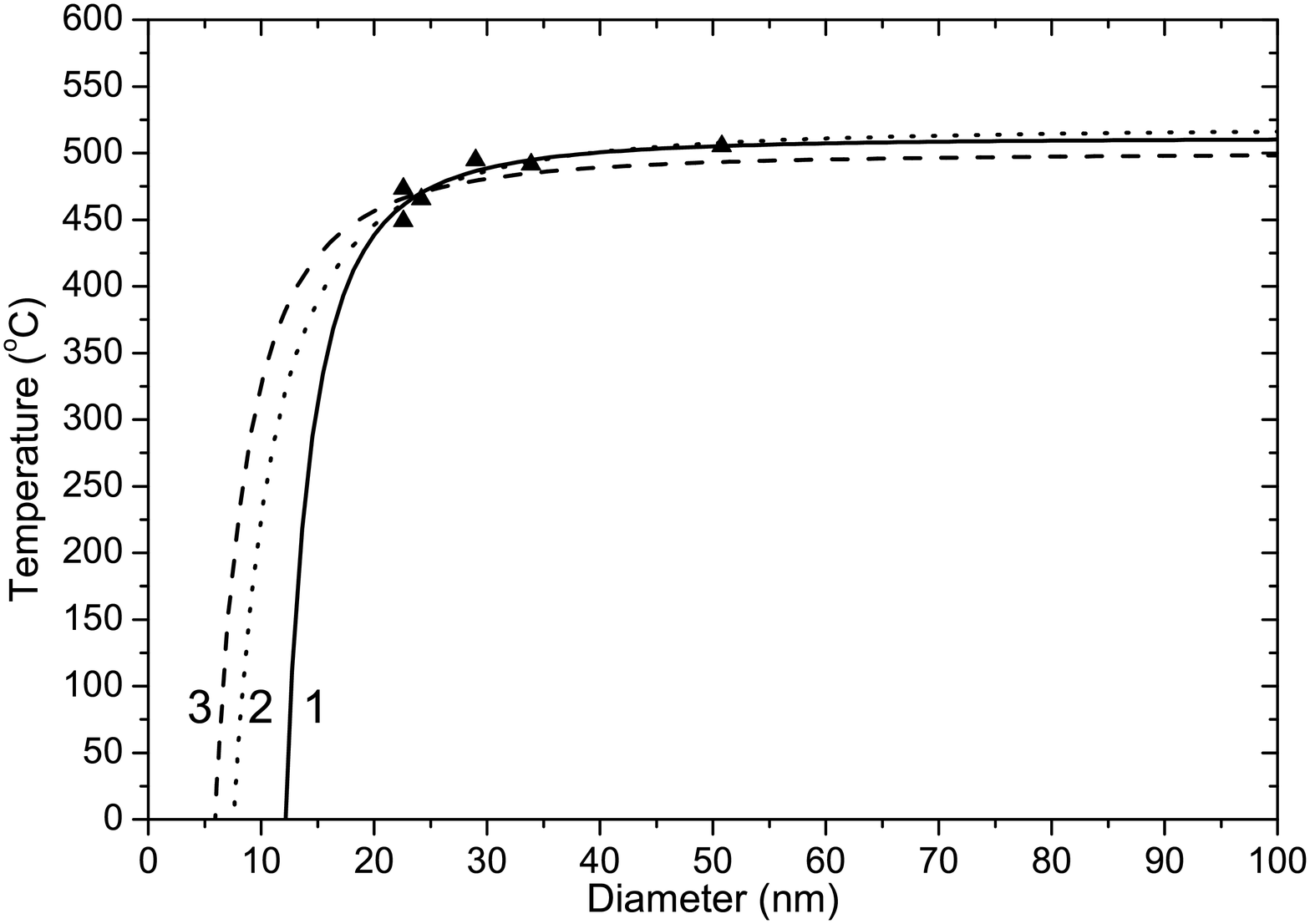}
    \caption{Critical temperature $T_{c}$ in  degrees Celsius (according to experimental points - full triangles) dependence on diameter of particles - $PbTiO_{3}$. Theoretical curves are: (1) for $T_{c, \infty}$, $d_{o}$ and $d_{m}$ as fitting parameters, (2) for $T_{c, \infty}$ and $d_{o}$ as fitting parameters with $d_{m}=0$ nm, (3) for $d_{o}$ as a fitting parameter with $T_{c, \infty}=500^{o} C$ and $d_{m}=0$ nm.}
    \label{fig:1}
\end{figure}

\begin{figure}[b]
    \centering
        \includegraphics[width=1.00\textwidth]{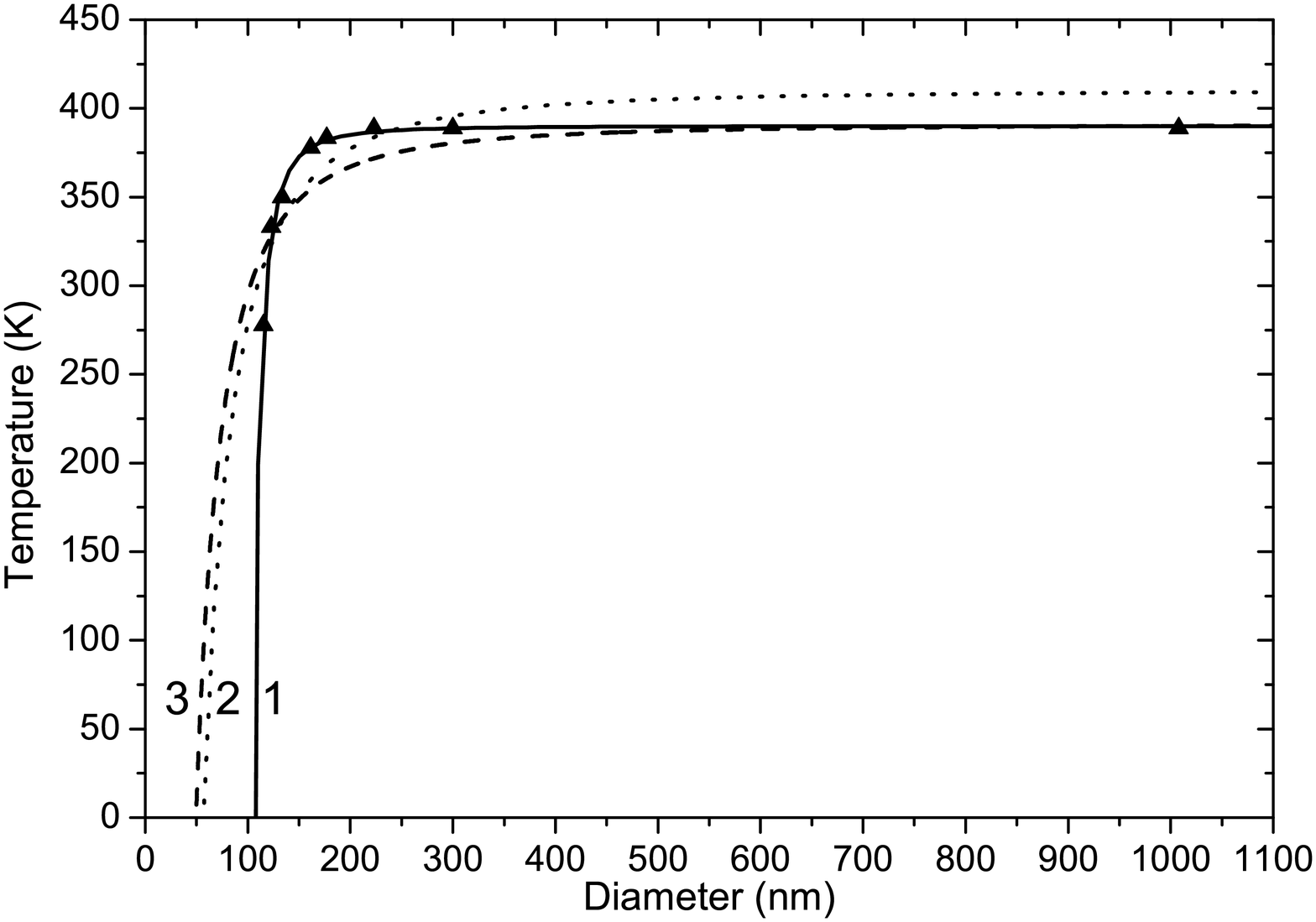}
    \caption{Critical temperature $T_{c}$ in Kelvins (according to experimental points - full traingles) dependence on diameter of particles - $BaTiO_{3}$. Theoretical curves are: (1) for $T_{c, \infty}$, $d_{o}$ and $d_{m}$ as fitting parameters, (2) for $T_{c, \infty}$ and $d_{o}$ as fitting parameters with $d_{m}=0$ nm, (3) for $d_{o}$ as a fitting parameter with $T_{c, \infty}=391 $ K and $d_{m}=0$ nm.}
    \label{fig:2}
\end{figure}

This formula is expected to correspond to the
Glinchuk et al. formula \cite{26} in some sense. In this formula, the
depolarisation and surface tension effects are taken into account for
diameters near the corresponding correlation diameter. Our above-given
results
were obtained using the exact calculation of the polarisation
profile induced by an external electric field in the paraelectric
phase for the large $\frac{D}{\delta}$ limit. The dependences from
(\ref{19}) with $\frac{1}{(d-d_{m})^{2}}$ and from (\ref{18})
$\frac{1}{d^{2}}$ are fitted to the experimental data for
$PbTiO_{3}$ , see Figure (\ref{fig:1}), and for $BaTiO_{3}$ , see
Figure (\ref{fig:2}). Far the best fit - the curve (1), is such in
which we are taking the critical temperature $T_{c, \infty}$, the
parameter $d_{o}$ and the parameter $d_{m}$ as fitting parameters
according to (\ref{19}). We have fitted this dependence (\ref{19})
also using the critical temperature as a fixed parameter taking
the values from the experiment and fitting the parameter $d_{o}$ and
the parameter $d_{m}$. We have found that such a fit is still
better than the fit using the formula (\ref{18}), see the curves
(2) and (3) respectively in Figure (\ref{fig:1}) and Figure
(\ref{fig:2}).

Critical temperature $T_{c}$ in degrees Celsius (we are using
degrees Celsius according to available experimental points)
dependence on diameter d of particles for $PbTiO_{3}$ is in Figure
(1). Theoretical curves are: curve (1) for $T_{c, \infty}$,
$d_{o}$ and $d_{m}$ as fitting parameters (we have found $T_{c,
\infty} = 511.5^{o}C$, $d_{o}=4.8$ nm and $d_{m}=7.3 $ nm with
quality of fitting procedure given by $R^{2}= 0.811$ and $d_{c}=
12.1 $ nm), curve (2) for $T_{c, \infty}$ and $d_{o}$ as fitting
parameters with $d_{m}=0 $ (we have found $T_{c, \infty} =
519.0^{o}$ C and $d_{o} = 7.5 $ nm with $R^{2}= 0.8024$ and
$d_{c}= 7.5 $ nm), curve (3) for $d_{o}$ as a fitting parameter
and $d_{m}=0$ (we have found $d_{o}=5.9 $ nm with $R^{2}= 0.6425 $
and $d_{c}= d_{o}$) with $T_{c, \infty}$ fixed to $500.0^{o}C$. In
\cite{16} regrarding $PbTiO_{3}$ ceramics the diameter corresponding to
our $d_{m}$ diameter is found to be 12.6 nm, compared to our dead
layer width 7.3 nm in Figure (\ref{fig:1}). The critical diameter
corresponding to our $d_{c} $ was not evaluated in \cite{16}. Note
that the dead layer diameter (=2 x dead layer radius) is in
\cite{16} interpreted as a critical size below which 
ferroelectricity becomes unstable.

Critical temperature $T_{c}$ in Kelvins (according to
available experimental points we are using Kelvins) dependence on
the diameter of particles - $BaTiO_{3}$. Theoretical curves are:
curve (1) for $T_{c, \infty}$, $d_{o}$ and $d_{m}$ as fitting
parameters (we have found $T_{c, \infty}=390.1$ K, $d_{o}=12.1 $
nm and $d_{m} = 92.8 $ nm with a quality of the fitting procedure given
by $R^{2}= 0.9938$ and with $d_{c}= 104.9 $ nm), curve (2) for
$T_{c, \infty}$ and $d_{o}$ as fitting parameters with $d_{m} = 0
$ (we have found $T_{c, \infty}= 410.2$ K and $d_{o}=56.5 $ nm
with $R^{2}= 0.7768 $ and $d_{c}= d_{o}$), curve (3) for $d_{o}$
as a fitting parameter with fixed $T_{c, \infty} = 391.0$ K from
\cite{21} and $d_{m}=0$ nm (we have found $d_{o} = 49.4 $ nm with
$R^{2}= 0.6940$ and $d_{c}= d_{o}$). In \cite{21} the
theoretically predicted value of the critical diameter is 44 nm
and experimentally found 115 nm, they should be compared to our
104.9 nm for the curve (1), 56.5 nm for the curve (2) and 49.4 nm
for the curve (3). We see that our value of the critical diameter
from the curve (1) is near the experimental value. Our values for
curves (2) and (3) are near the theoretical value (for which there
no dead layer was assumed in \cite{21}). In \cite{21.3} on dense
$BaTiO_{3}$ ceramics with grain size decreasing from 1200 to 50 nm,
the authors have found strong support for the existence of an
intrinsic size effect. From the experimental trends the critical
size for disappearance of ferroelectricity has been evaluated to
be from 10 nm to 30 nm. The strong depression of the relative
permittivity observed for the nanocrystalline ceramics can be
ascribed to the combination of the intrinsic size effect and of
the size-dependent dilution effect of a grain boundary dead layer.
We have found that the quantity $d_{m}$ has the value 92.8 nm for
the curve (1). This quantity, which contributes to the total
diameter of the particle, leads to the part of the diameter which
corresponds to polar region with d=11.9 nm for the critical
diameter $d_{c} = 104.9$ nm which we have found, see Figure
(\ref{fig:2}). Thus, if we identify $d_{m}$ with the dead layer
diameter, then we see that the critical diameter found by our
fitting procedure is 3 to 10 times larger than that found
experimentally.

\section{Conclusions}

A phase transition in a spheroidal particle from paraelectric to
ferroelectric phase as well as dynamic susceptibility are
affected by the spheroidal shape of ferroelectric particles.
This shape leads to many resonance frequences in the dynamic
susceptibility. Sources of the observed complicated
multirelaxation behaviour of the dynamic susceptibility in
ferroelectric-dielectric composites are of a different type, the
shape of particles in microcomposites is one of them. In our model
in this paper we assume that the surface charges are compensated.
In general, there is a different polarisation at the surface of
the particle from than in the bulk. The spheroidal shape does not lead
to plane polarisation waves due to the surface effect: the
boundary conditions in a spheroidal particle with a surface layer
lead to a multirelaxation response. We have found that for the
surface boundary condition $P=0$ on the surface there is an
infinite number of the responding modes as well as infinite
number of resonance frequencies. We have found formulas which
enable to discuss the effects of the ferroelectric particle shape
(spheroid) on its dielectric response. First we described a model
from which we have found the dynamic dielectric function within
the quasistatic approximation. Dynamic equations for the order
parameter, the electric polarisation vector component, were found.
We discussed the paraphase properties of the dielectric response of
the particle. The Cole-Cole diagram is not of the semicircle form.
The static susceptibility within the paraphase is discussed. It
does not have the Curie-Weiss like behaviour as in the bulk.

The transition temperature decreases with the decreasing diameter
d of the particle quadratically. We discussed our theoretical
results comparing them with experimental results for the critical
temperature diameter dependence in $PbTiO_{3}$ and in $BaTiO_{3}$.
We have found that the formula $T_{c}(d) = T_{c, \infty}(1 -
\frac{d_{o}^{2}}{(d-d_{m})^{2}})$ where $T_{c, \infty}$ is a
fitting parameter in general together with $d_{o}$ and $d_{m}$,
the last parameter is a parameter describing the dead layer
thickness.

\section*{Acknowledgement}

The author wishes to express his sincere thanks to V. Dvorak, J.
Petzelt and I. Rychetsky, the Department of Dielectrics, Institute of
Physics AS CR, Praha, for their discussions about microcomposite
materials and small particle ferroelectric properties, to D.
Nuzhnyy from the same Institute for his technical help with the
figures, to V. Trnovcova, Department of Physics, Faculty of
Materials Science and Technology, Trnava Slovak University of
Technology in Bratislava, Trnava, for discussions concerning the
microcomposites during the author's stay in Trnava and to V. Sepelak from the
Geotechnical Institute, SAS, Kosice for discussions on ferroics. This work was
supported partially by the grant VEGA 1/3042/06. This work was supported by the Slovak Research and Development Agency under the contract No. APVV-0728-07 at the final stages of the work. The author wishes
to express his sincere thanks to P. Presnajder from the Faculty of
Mathematics, Physics and Informatics, Comenius University, Bratislava, for his kind support.

\end{document}